\documentclass[prd,aps, preprintnumbers, showpacs,twocolumn, nofootinbib,superscriptaddress,notitlepage]{revtex4-1}
\usepackage[normalem]{ulem}
\usepackage{epsfig}
\usepackage{amsfonts}
\usepackage{amsmath}
\usepackage{slashed}
\usepackage{graphicx}
\usepackage{color}
\usepackage{mathtools}
\usepackage{subfigure}
\usepackage{graphicx}

\usepackage{stmaryrd}
\usepackage{amssymb,psfrag}
\usepackage[normalem]{ulem}
\usepackage{caption}

\newcommand{\beq}{\begin{eqnarray}}
\newcommand{\eeq}{\end{eqnarray}}

\newcommand{\nn}{\nonumber}

\def\slash#1{#1 \hskip-0.45em /}

\begin{document}

\title{Accessing the subleading-twist $B$-meson light-cone distribution amplitude with Large-Momentum Effective Theory}

\author{Shu-Man Hu}
\affiliation{School of Physics and Microelectronics, Zhengzhou University, Zhengzhou, Henan 450001, China}

\author{Wei Wang}
\email{Corresponding author. wei.wang@sjtu.edu.cn}
\affiliation{INPAC, Key Laboratory for Particle Astrophysics and Cosmology (MOE), Shanghai Key Laboratory for Particle Physics and Cosmology,
School of Physics and Astronomy, Shanghai Jiao Tong University, Shanghai 200240, China}

\author{Ji Xu}
\email{Corresponding author. xuji\_phy@zzu.edu.cn}
\affiliation{School of Physics and Microelectronics, Zhengzhou University, Zhengzhou, Henan 450001, China}

\author{Shuai Zhao}
\email{Corresponding author. zhaoshuai1986@gmail.com}
\affiliation{INPAC, Key Laboratory for Particle Astrophysics and Cosmology (MOE), Shanghai Key Laboratory for Particle Physics and Cosmology,
School of Physics and Astronomy, Shanghai Jiao Tong University, Shanghai 200240, China}

\begin{abstract}
  Within the framework of large momentum effective theory (LaMET), we propose a hard-collinear factorization formula to extract the subleading-twist $B$-meson light-cone distribution amplitude (LCDA)
  from the
  quasidistribution amplitude calculated on the lattice. The one-loop matching coefficient
  and an integro-differential equation governing the evolution of the quasidistribution amplitude are derived. Our results could be useful either in evaluating the subleading-twist $B$-meson LCDA on the lattice or in the understanding the feasibility of LaMET on higher-twist distributions.
\end{abstract}

\maketitle

\section{Introduction}
\label{Introduction}
The light-cone distribution amplitudes of $B$-meson, which are the inherent parts of  factorization theorems for many exclusive $B$ decay processes,  are of great phenomenological interest, and have been indispensable ingredients for precision calculations of the $B$-meson decay observables \cite{Grozin:1996pq,Beneke:2000wa,Beneke:1999br,Beneke:2001ev}. For example,  the LCDAs of $B$-meson are helpful in determining fundamental parameters in the flavor sector, such as the Cabibbo-Kobayashi-Maskawa (CKM) matrix, as well as the CP asymmetries in rare $B$ decays~\cite{Becher:2005bg,Na:2015kha,Gao:2021sav}. The semileptonic decays, e.g., $B \to \pi \ell \nu$ and $B \to D \ell \nu$, serve as the golden channels for the determinations of $|V_{ub}|$ and $|V_{cb}|$; however, the precision of theoretical prediction is limited by the uncertainties of the  LCDAs.

In HQET, the leading-twist LCDA of $B$-meson, which is denoted as  $\phi_B^+$,  gives a dominant contribution in the heavy-quark expansion, hence received considerable attentions \cite{Bell:2013tfa,Braun:2014owa,Feldmann:2014ika,Lu:2022kos}. The higher-twist distribution amplitudes give rise to power-suppressed contributions to $B$ decays in which the energetic light particles at final states exists. It is clear that the utility of QCD factorization theorem depends on the possibility to estimate the power corrections involving higher-twist DAs, and the accuracy is not sufficient when considering only the leading power contributions \cite{Braun:2017liq,Shen:2021yhe,Cheng:2014fwa}.

Besides, the investigation of higher-twist distribution amplitudes would be desirable and interesting by itself. For example, the subleading-twist LCDA $\phi_B^-(\omega,\mu)$ governs the leading-power contribution to $B\to D$ form factors \cite{Gao:2021sav}, and the knowledge of its anomalous dimension $\gamma_-$ is essential to check certain correlation functions for $B\to\pi$ form factors within soft-collinear effective theory (SCET) \cite{DeFazio:2005dx}.

The studies of the leading-twist $B$-meson LCDA $\phi_B^+(\omega,\mu)$ such as asymptotic behaviors, renormalization
group equation (RGE), equations of motion and etc., have been carried out for a relatively long time \cite{Kawamura:2001jm,Kawamura:2008vq,Braun:2019wyx,Wang:2018wfj,Beneke:2018wjp,Liu:2020ydl,Bell:2020qus}. On the other hand, the higher-twist DAs are attracting increasing attention lately. It was pointed out that the structure of subleading-twist DA is simpler than assumed~\cite{Bell:2008er,Braun:2015pha}. It evolves autonomously and does not mix with ``genuine'' three-particle contributions. However, despite these impressive achievements, our knowledge on LCDAs for $B$-meson is relatively poor. We have not been able to depict the leading and subleading-twist DAs since they encode information of the nonperturbative strong interaction dynamics from the soft-scale fluctuation, most of the recent studies on the shapes of $\phi_B^+(\omega,\mu)$ and $\phi_B^-(\omega,\mu)$ are model-dependent. Therefore,  a significant task in heavy flavor physics will be improving the accuracy of $B$-meson LCDAs  in a model-independent manner.

The LCDAs are presumably dominated by QCD interactions at low momentum scale so that they cannot be calculated with perturbation theory. One must resort to nonperturbative methods among which lattice QCD distinguishes itself due to being based on first principles and its rapid development in recent years. The large mass of the bottom quark makes it difficult to perform conventional lattice simulations, thus necessitates the use of effective field theories such as HQET in lattice simulations \cite{Lepage:1992tx,Blossier:2012qu,Brambilla:2017hcq,Bahr:2016ayy}. However, performing the lattice calculation of LCDAs directly is known to
be impractical by the appearance of nonlocal operators located on the light-cone.  The development of the ``lattice parton physics'', e.g.,  Large-Momentum Effective Theory (LaMET)~\cite{Ji:2013dva,Ji:2014gla} and related proposals such as pseudo-PDFs~\cite{Radyushkin:2017cyf,Radyushkin:2019mye} and lattice cross sections~\cite{Ma:2017pxb} provides a practical way to calculate distributions defined with light-cone operators on the lattice.
  The essential strategy of LaMET resides in the construction of an equal-time Euclidean quantity that can be computed on the lattice, and at the meantime, can be matched to the original LCDA by a factorization formula in the large momentum limit.  The past few years have witnessed fruitful results obtained in the frame of LaMET, which indicates a bright future to systematically compute a wide range of ``parton observables'' with satisfactory uncertainties. For the recent progresses, see reviews~\cite{Ji:2020ect,Zhao:2018fyu,Cichy:2018mum}  and references therein.

The attempt to apply LaMET to the leading-twist $B$-meson LCDA $\phi_B^+$ was presented in~\cite{Kawamura:2018gqz,Wang:2019msf}. The  $\phi_B^+$ was later explored within the pseudodistribution approach~\cite{Zhao:2020bsx}. In the followup  works the inverse moment and nonperturbative renormalization  of quasidistributions were discussed~\cite{Xu:2022krn,Xu:2022guw}.
It is natural to extend the previous works of $\phi_B^+$ to  $\phi_B^-$, which is the main goal of the present work.
Specifically, we will define the quasidistribution that corresponds to $\phi_B^-$,  derive the matching relation and the evolution of the quasidistribution amplitude, and explore the feasibilities for lattice calculations.

The remainder of this paper is organized as follows. In Sec.\,\ref{Subleading}, we present the definitions of the subleading-twist (twist-3) LCDA $\phi_B^-(\omega,\mu)$ and quasidistribution amplitude $\varphi_B^-(\xi,\mu)$. In Sec.\,\ref{CalculationandRenormalization}, we calculate the one-loop corrections of LCDA and quasidistribution amplitude as well as their renormalization equations. In Sec.\,\ref{FactorizationFormula}, the factorization formula and matching coefficient are exhibited, which are the main results of this work. In Sec.\,\ref{Perspectives}, we give a brief outlook for lattice calculations for $\phi_B^-(\omega,\mu)$ in terms of the results in this paper. The last section is the summary and perspective on future works.

\section{Subleading twist $B$-meson (quasi)Distribution amplitudes}
\label{Subleading}
The $B$-meson LCDAs in coordinate space are defined through the Lorentz decomposition of the renormalized matrix elements of the nonlocal operator which contains an effective heavy-quark field and a light-quark field separated on the light-cone
\begin{eqnarray}\label{decom1}
  && \left\langle 0\left|\bar{q}_\beta(\eta n_+) W(\eta n_+, 0) h_{v\alpha}(0)\right| \bar{B}(v)\right\rangle =  \nn\\
  &&\quad -\frac{ i \tilde f_B(\mu) M}{4}\bigg[\frac{1+\slash{v}}{2} \Big\{2 \tilde{\phi}^{+}_B(\eta,\mu)  \nn\\
  &&\quad + \Big(\tilde{\phi}^{-}_B(\eta,\mu) - \tilde{\phi}^{+}_B(\eta,\mu)\Big)\frac{ \slashed n_+}{  n_+ \!\cdot\! v  } \Big\} \gamma_5 \bigg]_{\alpha \beta} \,.
\end{eqnarray}
Where $W(\eta n_+, 0)=\mathrm{P}\left\{\operatorname{Exp}\left[i g_{s} \int_{0}^{\eta} d x \, n_+ \!\cdot\! A(x n_+)\right]\right\}$ is Wilson line connecting the light and heavy quark fields that ensures gauge invariance, $v_\mu$ is the heavy quark velocity satisfying $v^2=1$, $| \bar{B}(v) \rangle$ is the $\bar{B}$-meson state, $\tilde f_B(\mu)$ is the static decay constant in HQET \cite{Beneke:2005gs}. Here and below the notations for light-cone coordinates are
\begin{eqnarray}
	n_{+\mu}=\frac{1}{\sqrt{2}}(1,0,0,1) \,, \quad n_{-\mu}=\frac{1}{\sqrt{2}}(1,0,0,-1) \,.
\end{eqnarray}
The functions $\tilde{\phi}^{+}_B(\eta,\mu)$ and $\tilde{\phi}^{-}_B(\eta,\mu)$ are the two-particle leading- and subleading-twist $B$-meson LCDAs, respectively. The prefactor in Eq.\,(\ref{decom1}) is chosen in such a way that in case $\eta=0$, one obtains
\begin{eqnarray}\label{local1}
  \langle 0|\bar{q}(0) \gamma^\mu \gamma_5 h_{v}(0)| \bar{B}(v)\rangle = i \tilde f_B M v^\mu
\end{eqnarray}
with the normalization conditions
\begin{eqnarray}
  \tilde{\phi}^{+}_B(\eta=0,\mu)=\tilde{\phi}^{-}_B(\eta=0,\mu)=1 \,.
\end{eqnarray}
This indicates
\begin{eqnarray}\label{local2}
  \left\langle 0\left|\bar{q}(0) \slash{n}_+ \gamma_5 h_v(0)\right| \bar{B}(v)\right\rangle =  i \tilde f_B(\mu) M v^+ \,.
\end{eqnarray}

If we plug $\Gamma_{\beta\alpha} = [\slash{n}_+\gamma_5]_{\beta\alpha}$ into the matrix element of the nonlocal operator in Eq.\,(\ref{decom1}),
\begin{eqnarray}
  &&\left\langle 0\left|\bar{q}(\eta n_+) \slash{n}_+\gamma_5 W(\eta n_+, 0) h_v(0)\right| \bar{B}(v)\right\rangle \nn\\
  &&= i \tilde f_B(\mu) M \tilde\phi_B^+(\eta,\mu) v^+ \,.
\end{eqnarray}
For convenience, hereafter we subsequently denote $v^+ \equiv n_+\cdot v$ and $v^- \equiv n_-\cdot v$. Therefore we have
\begin{eqnarray}\label{tildephibt}
  && \tilde\phi_B^+(\eta,\mu) \nn\\
  &&\quad = \frac{1}{i\tilde f_B(\mu) M v^+} \left\langle 0\left|\bar{q}(\eta n_+) \slash{n}_+\gamma_5 W(\eta n_+, 0) h_v(0)\right| \bar{B}(v)\right\rangle \,. \nn\\
\end{eqnarray}
Applying the Fourier transformation for $\tilde\phi_B^+(\eta,\mu)$ leads to the momentum-space distribution amplitude, which usually appear in factorization formulas
\begin{eqnarray}\label{Fourier1}
  \phi_B^+(\omega,\mu) = \frac{v^+}{2\pi} \int_{-\infty}^{+\infty} d\eta \, e^{i\omega v^+ \eta} \, \tilde\phi_B^+(\eta,\mu)  \,.
\end{eqnarray}
Unlike in QCD, the LCDA defined in HQET depends on a dimensional argument $\omega$, it has the meaning of the light-cone projection of the light-quark momentum in the heavy-meson rest frame.

The leading-twist LCDA in momentum-space can be expressed as the ratio of the nonlocal and local matrix elements in Eq.\,(\ref{tildephibt}) and Eq.\,(\ref{local2}) respectively,
\begin{eqnarray}\label{defiofphi+}
  &&\phi_B^+(\omega,\mu) = v^+ \int_{-\infty}^{+\infty} \frac{d\eta}{2\pi} \, e^{i\omega v^+ \eta} \nn\\
   &&\quad \times  \frac{\left\langle 0\left|\bar{q}(\eta n_+) \slash{n}_+\gamma_5 W(\eta n_+, 0) h_v(0)\right| \bar{B}(v)\right\rangle}{\left\langle 0\left|\bar{q}(0) \slash{n}_+ \gamma_5 h_v(0)\right| \bar{B}(v)\right\rangle} \,.
\end{eqnarray}
Similarly, if $\Gamma_{\beta\alpha} = [\slash{n}_-\gamma_5]_{\beta\alpha}$, one has the subleading-twist LCDA
\begin{eqnarray}\label{defiofphi-}
  &&\phi_B^-(\omega,\mu) = v^+ \int_{-\infty}^{+\infty} \frac{d\eta}{2\pi} \, e^{i\omega v^+ \eta} \nn\\
   &&\quad \times \frac{\left\langle 0\left|\bar{q}(\eta n_+) \slash{n}_-\gamma_5 W(\eta n_+, 0) h_v(0)\right| \bar{B}(v)\right\rangle}{\left\langle 0\left|\bar{q}(0) \slash{n}_- \gamma_5 h_v(0)\right| \bar{B}(v)\right\rangle} \,.
\end{eqnarray}
Eqs.\,(\ref{defiofphi+}) and (\ref{defiofphi-}) provide expressions for the two-particle $B$-meson LCDAs. We will focus on the subleading-twist LCDA $\phi_B^-(\omega,\mu)$ in this work.

Next we need to identify a quasidistribution amplitude which is both calculable on the lattice and feasible for extracting the LCDA. Based on the construction idea in Ref.\,\cite{Ji:2013dva}, the following definition of $B$-meson quasidistribution amplitude is employed:
\begin{eqnarray}\label{defiofquasiphi-}
  &&\varphi_B^-(\xi,\mu) = v^z \int_{-\infty}^{+\infty} \frac{d\tau}{2\pi} \, e^{i\xi v^z \tau} \nn\\
   && \quad \times \frac{\left\langle 0\left|\bar{q}(\tau n_z) (\gamma^t-\gamma^z)\gamma_5 W(\tau n_z, 0) h_v(0)\right| \bar{B}(v)\right\rangle}{\left\langle 0\left|\bar{q}(0) (\gamma^t-\gamma^z)\gamma_5 h_v(0)\right| \bar{B}(v)\right\rangle} \,.\nn\\
\end{eqnarray}
It is readily seen that $\varphi_B^-(\xi,\mu)$ is constructed by the spatial correlation function of two (effective) quark fields with $n_z=(0,0,0,1)$ and $v^z \equiv n_z \cdot v$. We will work in a Lorentz boosted frame of the $B$-meson in which $v^+ \gg v^-$ and set $v_{\perp\mu}=0$. Unlike $\phi_B^-(\omega,\mu)$ in Eq.\,(\ref{defiofphi-}), which is invariant under a boost along the $z$ direction, the quasidistribution amplitude changes dynamically under such a boost, which is encoded in its nontrivial dependence on the heavy quark velocity.

\section{One-loop Calculation and Renormalization group equation}
\label{CalculationandRenormalization}

\subsection{Preliminaries}
To examine the factorization formula and determine the matching coefficient, we replace the $B$-meson state with a heavy $b$ quark plus an off-shell light-quark and perturbatively calculate the radiative corrections of subleading-twist LCDA and quasidistribution amplitude in this section. We carry out the calculation by utilizing the off-shellness of the light-quark as an IR regulator; and the dimensional regularization with modified minimum subtraction scheme ($\overline{\textrm{MS}}$) is implied.

\begin{figure}[htbp]
	\centering
	\includegraphics[width=0.8\columnwidth]{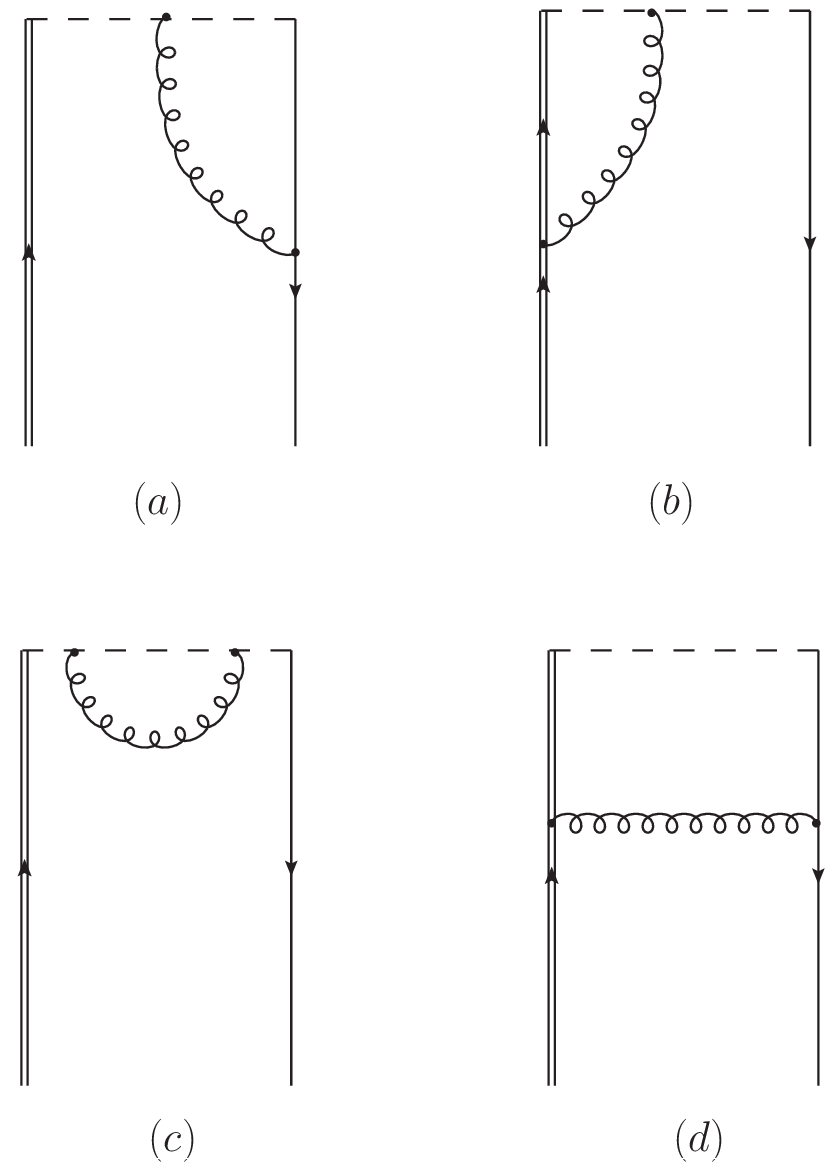}
	\caption{Feynman diagrams for the LCDA and quasidistribution amplitude of $B$-meson. The effective HQET  bottom quark is represented by the double line; the Wilson line is indicated by the dashed line.}
	\label{All-in}
\end{figure}

We take the off-shellness of the initial light-quark
\begin{eqnarray}
  k^2 = 2k^+ k^- - k_\perp^2
\end{eqnarray}
and set the space-time dimension $d=4-2\epsilon$. The Feynman diagrams at the one-loop level are shown in Fig.\,\ref{All-in}. The distribution amplitudes of Fock state can be expanded in series of $\alpha_s$.

Take the subleading-twist LCDA for instance, define
\begin{eqnarray}\label{defofOperatorsLCDA}
  \left\{\begin{array}{ll}
  O_+(\eta) = \bar{q}(\eta n_+) \slash{n}_-\gamma_5 W(\eta n_+, 0) h_v(0) \,,\\
  O_+(0) = \bar{q}(0) \slash{n}_- \gamma_5 h_v(0) \,.
  \end{array} \right.
\end{eqnarray}
Therefore up to the one-loop level, $\phi_B^-(\omega,\mu)$ in Eq.\,(\ref{defiofphi-}) can be expressed as
\begin{eqnarray}\label{expanofphiB-1}
  &&\phi_B^-(\omega,\mu) = v^+ \int_{-\infty}^{+\infty} \frac{d\eta}{2\pi} \, e^{i\omega v^+ \eta} \nn\\
   &&\qquad \times \frac{ \left\langle 0\left|O_+(\eta)\right| b\bar q(k)\right\rangle^{(0)} +\left\langle 0\left|O_+(\eta)\right| b\bar q(k)\right\rangle^{(1)} }{ \left\langle 0\left|O_+(0)\right| b\bar q(k)\right\rangle^{(0)} +\left\langle 0\left|O_+(0)\right| b\bar q(k)\right\rangle^{(1)} } \nn\\
  && \quad = v^+ \int_{-\infty}^{+\infty} \frac{d\eta}{2\pi} \, e^{i\omega v^+ \eta} \nn\\
   &&\qquad \times \left[ \frac{\left\langle 0\left|O_+(\eta)\right| b\bar q(k)\right\rangle^{(0)}}{\left\langle 0\left|O_+(0)\right| b\bar q(k)\right\rangle^{(0)}} +\frac{\left\langle 0\left|O_+(\eta)\right| b\bar q(k)\right\rangle^{(1)}}{\left\langle 0\left|O_+(0)\right| b\bar q(k)\right\rangle^{(0)}} \right. \nn\\
  &&\qquad \left.-\frac{\left\langle 0\left|O_+(\eta)\right| b\bar q(k)\right\rangle^{(0)}}{\left\langle 0\left|O_+(0)\right| b\bar q(k)\right\rangle^{(0)}}\frac{\left\langle 0\left|O_+(0)\right| b\bar q(k)\right\rangle^{(1)}}{\left\langle 0\left|O_+(0)\right| b\bar q(k)\right\rangle^{(0)}} \right] +\mathcal{O}(\alpha_s^2) \,.\nn\\
\end{eqnarray}
Here the superscripts $(0)$ indicates the result at tree level and $(1)$ denotes the one-loop correction.

On the other hand, $\phi_B^-(\omega,\mu)$ can be organized as follows
\begin{eqnarray}\label{expanofphiB-2}
  \phi_B^-(\omega,\mu) = \phi_B^{-(0)}(\omega) + \phi_B^{-(1)}(\omega,\mu) + \mathcal{O}(\alpha_s^2) \,.
\end{eqnarray}
By comparing Eqs.\,(\ref{expanofphiB-1}) and (\ref{expanofphiB-2}), one can identify that
\begin{eqnarray}
  && \phi_B^{-(0)}(\omega) = v^+ \int_{-\infty}^{+\infty} \frac{d\eta}{2\pi} \, e^{i\omega v^+ \eta} \frac{\left\langle 0\left|O_+(\eta)\right| b\bar q(k)\right\rangle^{(0)}}{\left\langle 0\left|O_+(0)\right| b\bar q(k)\right\rangle^{(0)}} \nn\\
  &&\qquad \qquad = \delta(\omega-\tilde k) \,,\\ \nn\\
  &&\phi_B^{-(1)}(\omega,\mu) = v^+ \int_{-\infty}^{+\infty} \frac{d\eta}{2\pi} \, e^{i\omega v^+ \eta} \nn\\
   &&\quad \times \left[ \frac{\left\langle 0\left|O_+(\eta)\right| b\bar q(k)\right\rangle^{(1)}}{\left\langle 0\left|O_+(0)\right| b\bar q(k)\right\rangle^{(0)}} \right. \nn\\
  &&\quad \left.-\frac{\left\langle 0\left|O_+(\eta)\right| b\bar q(k)\right\rangle^{(0)}}{\left\langle 0\left|O_+(0)\right| b\bar q(k)\right\rangle^{(0)}}\frac{\left\langle 0\left|O_+(0)\right| b\bar q(k)\right\rangle^{(1)}}{\left\langle 0\left|O_+(0)\right| b\bar q(k)\right\rangle^{(0)}} \right] \,,
\end{eqnarray}
where $\tilde k = k^+/v^+ = k^z/v^z$. The general discussions of one-loop correction above are also applicable to the quasidistribution amplitude. In analogy with Eq.\,(\ref{defofOperatorsLCDA}), define
\begin{eqnarray}\label{defofOperatorsQuasi}
  \left\{\begin{array}{ll}
  O_z(\tau) = \bar{q}(\tau n_z) (\gamma^t-\gamma^z)\gamma_5 W(\tau n_z, 0) h_v(0) \,,\\
  O_z(0) = \bar{q}(0) (\gamma^t-\gamma^z)\gamma_5 h_v(0) \,.
  \end{array} \right.
\end{eqnarray}
Therefore, we have
\begin{eqnarray}
  && \varphi_B^{-(0)}(\xi) = v^z \int_{-\infty}^{+\infty} \frac{d\tau}{2\pi} \, e^{i\xi v^z \tau} \frac{\left\langle 0\left| O_z(\tau)\right| b\bar q(k)\right\rangle^{(0)}}{\left\langle 0\left|O_z(0)\right| b\bar q(k)\right\rangle^{(0)}} \nn\\
  &&\qquad \qquad = \delta(\xi-\tilde k) \,,\\ \nn\\
  &&\varphi_B^{-(1)}(\xi,\mu) = v^z \int_{-\infty}^{+\infty} \frac{d\tau}{2\pi} \, e^{i\xi v^z \tau} \nn\\
   &&\quad \times \left[ \frac{\left\langle 0\left| O_z(\tau)\right| b\bar q(k)\right\rangle^{(1)}}{\left\langle 0\left| O_z(0)\right| b\bar q(k)\right\rangle^{(0)}} \right. \nn\\
  &&\quad \left.-\frac{\left\langle 0\left| O_z(\tau)\right| b\bar q(k)\right\rangle^{(0)}}{\left\langle 0\left| O_z(0)\right| b\bar q(k)\right\rangle^{(0)}}\frac{\left\langle 0\left| O_z(0)\right| b\bar q(k)\right\rangle^{(1)}}{\left\langle 0\left| O_z(0)\right| b\bar q(k)\right\rangle^{(0)}} \right] \,.
\end{eqnarray}

\begin{widetext}
\subsection{One-loop results for distribution amplitudes}

\verb""We now list the results of one-loop corrections for subleading-twist LCDA and quasidistribution amplitude diagram by diagram. First, consider the amplitude of the light-quark sail diagram for LCDA in Fig.\,\ref{All-in}(a),
\begin{align}\label{LCDAa}
  \phi_{B}^{-(a)}(\omega,\mu)
  =&\frac{\alpha_sC_F}{2\pi}\left\{\begin{aligned}
    & \left[
    -\frac{1}{\epsilon}\frac{\omega}{(\omega-\tilde k)\tilde k} -\frac{\omega}{(\omega-\tilde k)\tilde k}\ln\frac{\mu^2\tilde k^{2}}{\omega(\tilde k-\omega)(-k^2)}\right]_\oplus  &0<\omega<\tilde k  \\
    & 0  & \textrm{Others} \\
  \end{aligned}\right. \nn\\
  -& \frac{\alpha_sC_F}{2\pi} \Bigg[ \left( \ln2-1 \right)\bigg(\frac{1}\epsilon+\ln\frac{\mu^2}{-k^2} \bigg) +\ln^2 2 -2+\frac{\pi^2}{12} \Bigg]\delta(\omega-\tilde k) \,.
\end{align}
The plus distribution is defined by
\begin{eqnarray}\label{defplus}
\left \{ {F}(\omega, \tilde k) \right \}_{\oplus} =  {F}(\omega, \tilde k)
- \delta(\omega-\tilde k) \, \int_0^{2 \omega}  \, d t \,  {F}(\omega, t) \,,
\end{eqnarray}
An advantage of introducing this plus function is that it allows to implement both the ultraviolet and infrared subtractions for the perturbative matching procedure simultaneously.

The contribution of Fig.\,\ref{All-in}(a) to quasidistribution amplitude is
\begin{align}\label{quasiDAa}
  \varphi_{B}^{-(a)}(\xi,\mu)
  =&\frac{\alpha_sC_F}{4\pi}\left\{\begin{aligned}
    & \left[\frac{1}{\tilde k(\xi-\tilde k)} \left( -\tilde k +2\xi\ln\frac{-\xi}{\tilde k-\xi} \right) \right]  & \xi<0  \\
    & \left[\frac{1}{\tilde k(\xi-\tilde k)} \left( 2\xi -\tilde k -2\xi\ln\frac{4\tilde k^2 v^{z2}}{-k^2} \right) \right]_\oplus  & 0<\xi<\tilde k  \\
    & \left[\frac{1}{\tilde k(\xi-\tilde k)} \left( \tilde k -2\xi\ln\frac{\xi}{\xi-\tilde k} \right) \right]_\oplus  &\xi>\tilde k  \\
  \end{aligned}\right. \nn\\
  + & \frac{\alpha_sC_F}{4\pi} \Bigg[ \frac{1}{\epsilon} +\ln\frac{\mu^2}{v^{z2}\xi^2} +2(1-\ln2)\ln\frac{4\tilde k^2 v^{z2}}{-k^2} +2-\frac{\pi^2}{3} \Bigg]\delta(\xi-\tilde k) \,.
\end{align}
Here, we assign $v^{\mu} = \left(v^0, 0, 0, v^z\right)$ with $v^2=1$ and $v^z\gg 1$.

Second, we list the amplitude of the heavy-quark sail diagram for LCDA in Fig.\,\ref{All-in}(b),
\begin{align}\label{LCDAb}
  \phi_{B}^{-(b)}(\omega,\mu)
  =&\frac{\alpha_sC_F}{2\pi}\left\{\begin{aligned}
    & \left[ \frac{1}{\epsilon} \frac{1}{\omega-\tilde k} +\frac{1}{\omega-\tilde k}\ln\frac{\mu^2}{(\omega-\tilde k)^2} \right]_\oplus  & \omega>\tilde k  \\
    & 0  & \textrm{Others} \\
  \end{aligned}\right. \nn\\
  -& \frac{\alpha_sC_F}{2\pi} \Bigg[ \frac{1}{2\epsilon^2} +\frac{1}{2\epsilon}\ln\frac{\mu^2}{\omega^2}  +\frac14\ln^2\frac{\mu^2} {\omega^2} +\frac{\pi^2}{24}\Bigg]\delta(\omega-\tilde k) \,.
\end{align}
It is worth noting that the dashed line in Fig.\,\ref{All-in}(b) is a lightlike Wilson line; the heavy-quark propagator can be regarded as a timelike Wilson line. The vertex correction to a timelike-lightlike cusp on Wilson line is known to have $1/\epsilon^2$ UV divergence at one loop, as shown in Eq.\,(\ref{LCDAb}). This is different from the case in light-meson LCDAs, since the heavy $b$-quark here is treated in HQET which modifies the divergence structure of the loop integrals.

The contribution of Fig.\,\ref{All-in}(b) to quasidistribution amplitude is
\begin{align}\label{quasiDAb}
  \varphi_{B}^{-(b)}(\xi,\mu)
  =&\frac{\alpha_sC_F}{4\pi} \left\{\begin{aligned}
    & \left[ \frac{1}{\xi-\tilde k} \left( \frac{1}{\epsilon} -\ln4v^{z2} +\ln\frac{\mu^2}{(\tilde k-\xi)^2} \right) \right]_\oplus  & \xi<\tilde k  \\
    & \left[ \frac{1}{\xi-\tilde k} \left( \frac{1}{\epsilon} +\ln4v^{z2} +\ln\frac{\mu^2}{(\xi-\tilde k)^2} \right) \right]_\oplus  &\xi>\tilde k  \\
  \end{aligned}\right. \nn\\
   -& \frac{\alpha_sC_F}{4\pi} \Bigg[ \bigg(\frac{1}{\epsilon}+\ln\frac{\mu^2}{4v^{z2}\xi^2}\bigg)\ln4v^{z2} +\frac12\ln^2 4v^{z2} -\frac{\pi^2}{6} \Bigg]\delta(\xi-\tilde k) \,.
\end{align}

Fig.\,\ref{All-in}(c) is the one-loop correction to the Wilson line's self energy, which is determined by the direction of the Wilson line. This amplitude for LCDA is proportion to $n_+^2=0$, thus gives zero contribution. For quasidistribution amplitude, it reads
\begin{align}\label{quasiDAc}
  \varphi_{B}^{-(c)}(\xi,\mu)
  =&\frac{\alpha_sC_F}{2\pi} \left\{\begin{aligned}
    & \left[ \frac{1}{\tilde k-\xi} \right]_\oplus  & \xi<\tilde k  \\
    & \left[ \frac{1}{\xi-\tilde k} \right]_\oplus  &\xi>\tilde k  \\
  \end{aligned}\right. \nn\\
   -& \frac{\alpha_sC_F}{2\pi} \Bigg[ \frac{1}{\epsilon} +\ln\frac{\mu^2}{4v^{z2}\xi^2} +2\Bigg]\delta(\xi-\tilde k) \,.
\end{align}

Finally, the contribution of box diagram in Fig.\,\ref{All-in}(d) for LCDA is
\begin{align}\label{LCDAd}
  \phi_{B}^{-(d)}(\omega,\mu)
  =&\frac{\alpha_sC_F}{2\pi}\left\{\begin{aligned}
    & \left[ \frac{1}{\epsilon}\frac{1}{\tilde k} +\frac{\omega}{\tilde k(\tilde k-\omega)}\left( 1-\ln\frac{\tilde k^2}{-k^2} \right) +\frac{1}{\tilde k}\ln\frac{\mu^2}{(\tilde k-\omega)\omega} \right]_\oplus  &0<\omega<\tilde k  \\
    & \left[ \frac{1}{\tilde k-\omega} \left( 1-\ln\frac{\omega}{\omega-\tilde k} \right) \right]_\oplus  &\omega>\tilde k  \\
    & 0  & \textrm{Others} \\
  \end{aligned}\right. \nn\\
  -& \frac{\alpha_sC_F}{2\pi} \Bigg[   (1-\ln2)\bigg(\frac{1}{\epsilon}+\ln\frac{\mu^2}{-k^2}\bigg)+1 -\operatorname{Li}_2(-2) -\frac32\ln^2 2 +\frac{\pi^2}{12}\Bigg]\delta(\omega-\tilde k) \,.
\end{align}
Notice that there is UV-divergent piece in Eq.\,(\ref{LCDAd}), which is different from the situation in leading-twist LCDA. This would result in a different anomalous dimension for subleading-twist LCDA, we will discuss this issue later.

For quasidistribution amplitude, Fig.\,\ref{All-in}(d) yields
\begin{align}\label{quasiDAd}
  \varphi_{B}^{-(d)}(\xi,\mu)
  =&\frac{\alpha_sC_F}{2\pi} \left\{\begin{aligned}
    & \left[ \frac{1}{\tilde k}\ln\frac{\tilde k -\xi}{-\xi} \right]_\oplus  & \xi<0  \\
    & \left[ \frac{1}{\tilde k(\tilde k-\xi)} \left( \xi(1-\ln4)+\tilde k\ln(4v^{z2}) -\xi\ln\frac{\tilde k^2 v^{z2}}{-k^2} \right) \right]_\oplus  & 0<\xi<\tilde k  \\
    & \left[ \frac{1}{\tilde k(\tilde k-\xi)} \left( \tilde k -\xi\ln\frac{\xi}{\xi-\tilde k} \right) \right]_\oplus  &\xi>\tilde k  \\
  \end{aligned}\right. \nn\\
   +& \frac{\alpha_sC_F}{2\pi} \Bigg[ (\ln2-1)\ln\frac{4\tilde k^2 v^{z2}}{-k^2} +1+ \textrm{Li}_2(-2) +\frac{\ln^2 2}{2}-3\ln3+4\ln2  +\frac{\pi^2}{6}\Bigg]\delta(\xi-\tilde k) \,.
\end{align}

In the above results, the subleading-twist LCDA is nonzero only in the physical region $\omega>0$, while the quasidistribution amplitude has nonzero support in all region $-\infty<\xi<+\infty$. Recall that we have used the off-shellness of light-quark $-k^2$ as IR regulator, this logarithmic IR singularities would cancel between quasidistribution amplitude and LCDA, leaving the matching coefficient independent on $-k^2$, as it should be. We will see this point clearly in the next section.
\end{widetext}

\subsection{Renormalization group equation}
Having the one-loop results in Eqs.\,(\ref{LCDAa})-(\ref{quasiDAd}), we can derive integro-differential equations governing the evolution of subleading-twist LCDA and quasidistribution amplitude. One of the reasons for the importance of renormalization group equation (RGE) is that it gives insight in the expected behavior of the DAs at large and small momenta, which is important for the status of factorization theorems for $B$-meson decays. Besides, a thorough understanding of scale dependence of quasidistribution amplitude would help its future simulations on the lattice.

We take $O_+(\omega)$ to denote the Fourier transformation of the bilocal HQET operator $O_+(\eta)$ in Eq.\,(\ref{defofOperatorsLCDA}) and write down the relation between bare and renormalized operators as follows
\begin{eqnarray}
  O_{-}^{\mathrm{ren}}(\omega, \mu)=\int d \omega' Z_{-}(\omega, \omega', \mu) \, O_{-}^{\mathrm{bare}}\left(\omega'\right) \,,
\end{eqnarray}
where $Z_{-}^{(0)} (\omega, \omega') = \delta(\omega-\omega')$ at lowest order. In the $\overline{\textrm{MS}}$ scheme, the renormalization constant $Z_{-}(\omega, \omega', \mu)$ is defined so as to subtract the ultraviolet divergences in the matrix element of the bare operator. The subleading-twist LCDA obeys the evolution equation
\begin{eqnarray}\label{RGEofphi-}
  \frac{d}{d \ln \mu} \phi_B^- (\omega, \mu) = - \int_0^{\infty} d \omega' \gamma_{-} (\omega, \omega', \mu) \, \phi_B^- (\omega', \mu) \,,\nn\\
\end{eqnarray}
with the anomalous dimension
\begin{eqnarray}\label{RGEofphi-2}
   && \gamma_-(\omega, \omega', \mu) = \nn\\
   && \quad - \int d\tilde \omega \, \frac{dZ_-(\omega,\tilde \omega, \mu)}{d\ln\mu} \, Z_-^{-1}(\tilde\omega, \omega', \mu) -\gamma_F \, \delta(\omega-\omega') \,.\nn\\
\end{eqnarray}
Here $\gamma_F$ is the anomalous dimension of the static decay constant $\tilde f_B(\mu)$ in HQET.

One can tell from Eq.\,(\ref{RGEofphi-}) that the renormalization properties of $\phi_B^- (\omega, \mu)$ is similar to the leading-twist LCDA. $\gamma_-(\omega, \omega', \mu)$ can be read off the UV-divergent terms in Eqs.\,(\ref{LCDAa}), (\ref{LCDAb}) and (\ref{LCDAd}):
\begin{eqnarray}\label{expofgamma-}
&& \gamma_{-}\left(\omega, \omega^{\prime} , \mu\right) = \frac{\alpha_sC_F}{4\pi}\Bigg\{ \left(4 \ln \frac{\mu}{\omega}-2\right) \delta\left(\omega-\omega^{\prime}\right) \nn\\
&& \quad -4 \omega\left[\frac{\theta\left(\omega^{\prime}-\omega\right)}{\omega^{\prime}\left(\omega^{\prime}-\omega\right)}
+\frac{\theta\left(\omega-\omega^{\prime}\right)}{\omega\left(\omega-\omega^{\prime}\right)}\right]_{+}  -4 \frac{\theta\left(\omega^{\prime}-\omega\right)}{\omega^{\prime}} \Bigg\} \,. \nn\\
\end{eqnarray}
This result repeats the RGE derived in Ref.\,\cite{Bell:2008er}.

The case of quasidistribution amplitude is analogous to the subleading-twist LCDA, we now write down the relation between bare and renormalized operators
\begin{eqnarray}
  O_{z}^{\mathrm{ren}}(\xi, \mu)=\int d \xi' Z_{z}(\xi, \xi', \mu) \, O_{z}^{\mathrm{bare}}\left(\xi'\right) \,.
\end{eqnarray}
The renormalization constant $Z_{z}(\xi, \xi', \mu)$ can be expanded in series of $\alpha_s$
\begin{eqnarray}
  Z_{z}(\xi, \xi', \mu) &=& Z_{z}^{(0)} (\xi, \xi') + Z_{z}^{(1)} (\xi, \xi', \mu) + \mathcal{O}(\alpha_s^2) \,,
\end{eqnarray}
here $Z_{z}^{(0)} (\xi, \xi') = \delta(\xi-\xi')$ at lowest order. According to the UV-divergent terms in Eqs.\,(\ref{quasiDAa}), (\ref{quasiDAb}), (\ref{quasiDAc}) and (\ref{quasiDAd}), we obtain
\begin{eqnarray}\label{Zz1}
  && Z_{z}^{(1)} (\xi, \xi', \mu) = - \frac{\alpha_sC_F}{4\pi} \bigg\{ \frac{1}{\epsilon} \left[ \frac{\theta(\xi-\xi')}{\xi-\xi'} - \frac{\theta(\xi'-\xi)}{\xi'-\xi} \right]_{\oplus} \nn\\
  && \quad -\left( \frac{1}{2\epsilon} +\frac{1}{\epsilon} \ln 4v^{z2} \right) \delta(\xi-\xi') \bigg\} \,.
\end{eqnarray}
The quasidistribution amplitude obeys the evolution equation
\begin{eqnarray}\label{RGEofvarphi}
  \frac{d}{d \ln \mu} \varphi_B^- (\xi, \mu) = - \int_0^{\infty} d \xi' \gamma_{z} (\xi, \xi', \mu) \, \varphi_B^- (\xi', \mu) \,,
\end{eqnarray}
with the anomalous dimension
\begin{eqnarray}\label{RGEofvarphi2}
   && \gamma_z(\xi, \xi', \mu) = \nn\\
   &&\quad - \int d\tilde \xi \, \frac{dZ_z(\xi,\tilde \xi, \mu)}{d\ln\mu} \, Z_z^{-1}(\tilde\xi, \xi', \mu) -\gamma_F \, \delta(\xi-\xi') \,.\nn\\
\end{eqnarray}
Substituting Eq.\,(\ref{Zz1}) into Eq.\,(\ref{RGEofvarphi2}), one has
\begin{eqnarray}\label{expofgammaz}
&& \gamma_{z}\left(\xi, \xi', \mu\right) = \frac{\alpha_sC_F}{4\pi}\Bigg\{ \left( 4 +2\ln 4v^{z2} \right) \delta\left(\xi-\xi'\right) \nn\\
&& \quad -2 \left[ \frac{\theta(\xi-\xi')}{\xi-\xi'} - \frac{\theta(\xi'-\xi)}{\xi'-\xi} \right]_{\oplus}   \Bigg\} \,.
\end{eqnarray}
The result in Eq.\,(\ref{expofgammaz}) gives the scale behavior of the quasidistribution amplitude. As we have mentioned before, the quasidistribution amplitude changes dynamically under the boost of heavy quark velocity.

\section{Hard-collinear Factorization Formula}
\label{FactorizationFormula}
We now proceed to determine the perturbative matching coefficient entering the hard-collinear factorization formula for quasidistribution amplitude. Following the construction in Ref.\,\cite{Wang:2019msf}, the factorization formula is
\begin{eqnarray}\label{facformula}
	\varphi_B^-(\xi,\mu) = \int_{0}^{\infty} d \omega H\left(\xi, \omega, v^z, \mu\right) \phi_{B}^{-}(\omega, \mu)
    +\mathcal{O}\left(\frac{\Lambda_{\mathrm{QCD}}}{v^z \xi}\right) \,.\nn\\
\end{eqnarray}
With the spirit of LaMET, one can expect the IR physics of quasidistribution and subleading-twist LCDA are the same. The difference in the UV behavior between these two quantities is denoted by the matching coefficient $H$. Because QCD is asymptotic free, this difference is perturbatively calculable, as only the high-momentum modes matter. This property makes it possible to extract light-cone parton physics from quasiquantities.

Based on the calculations in Sec.\,\ref{CalculationandRenormalization} and the factorization formula in Eq.\,(\ref{facformula}), the matching coefficient $H$ is then determined by the difference between the momentum-space quasidistribution amplitude and the subleading-twist LCDA. We expand them in series of $\alpha_s$ up to the one-loop level:
\begin{eqnarray}
	&& \varphi^-_B(\xi,\mu) = \varphi^{-(0)}_B(\xi) + \varphi^{-(1)}_B(\xi,\mu) + \mathcal{O}(\alpha^2) \,,\nn\\
	&& \phi_B^-(\omega,\mu) = \phi_B^{-(0)}(\omega) + \phi_B^{-(1)}(\omega,\mu) + \mathcal{O}(\alpha^2) \,,\nn\\
	&& H(\xi, \omega, v^z, \mu) = H^{(0)}(\xi, \omega) + H^{(1)}(\xi, \omega, v^z, \mu) \nn\\
    && \qquad\qquad\qquad\quad + \mathcal{O}(\alpha^2) \,.
\end{eqnarray}
With the tree-level results of $\phi_B^-$ and $\varphi_B^-$ one can immediately get
\begin{eqnarray}
   && H^{(0)}(\xi, \omega)  = \delta(\xi-\omega) \,.
\end{eqnarray}
Substitute the expressions above into Eq.\,(\ref{facformula}), the one-loop result of matching coefficient can be expressed as
\begin{eqnarray}\label{calofH1}
	\left. H^{(1)}(\xi, \omega, v^z, \mu)\right|_{\omega\to \tilde k} = \varphi^{-(1)}_B(\xi, \mu)- \left. \phi_B^{-(1)}(\omega,\mu)\right|_{\omega\to\xi} \,. \nn\\
\end{eqnarray}
With the one-loop correction $\varphi^{-(1)}_B(\xi, \mu)$ and $\phi_B^{-(1)}(\omega,\mu)$  calculated in Sec.~\ref{CalculationandRenormalization},  one can obtain the hard matching coefficient
\begin{widetext}
\begin{eqnarray}\label{resultofH}
  H\left(\xi, \omega, v^z, \mu\right) &=& \delta(\xi-\omega) + \frac{\alpha_s C_F}{4\pi} \Bigg\{ \left[ \frac{1}{\omega-\xi}\left( 3-\ln\frac{\mu^2}{4v^{z2}(\xi-\omega)^2} -\ln\frac{\xi^2}{(\xi-\omega)^2} \right) \right] \theta(-\xi)\theta(\omega) \nn\\
  &&+\left[ \frac{1}{\omega(\omega-\xi)}\left( 3\omega-2\xi-3\omega\ln\frac{\mu^2} {4 v^{z2} (\omega-\xi)^2}+ \omega\ln \frac{\xi^2}{(\omega-\xi)^2}\right) \right]_{\oplus} \theta(\xi)\theta(\omega-\xi) \nn\\
  &&+\left[ \frac{1}{\xi-\omega}\left( -3+\ln\frac{\mu^2 }{4v^{z2}(\xi-\omega)^2}+\ln\frac{\xi^2}{(\xi-\omega)^2} \right) \right]_{\oplus} \theta(\omega)\theta(\xi-\omega) \nn\\
  &&+\left[\frac12\ln^2\frac{\mu^2}{4v^{z2}\xi^2}-\ln\frac{\mu^2}{4v^{z2}\xi^2}-2 -6\ln 3+ 5 \ln 4 +\frac{7\pi^2}{12}  \right] \delta(\xi-\omega)   \Bigg\} \,.
\end{eqnarray}
\end{widetext}
The logarithmic IR singularities $\ln(-k^2)$ cancel between the quasidistribution amplitude and the subleading-twist LCDA, leaving the obtained $H$ independent on the IR regulator as expected. Eq.\,(\ref{resultofH}) presents one of the main results of this work.

\section{Perspectives for Lattice Calculations}
\label{Perspectives}
\begin{figure}[htbp]
\centering
\includegraphics[width=0.9\columnwidth]{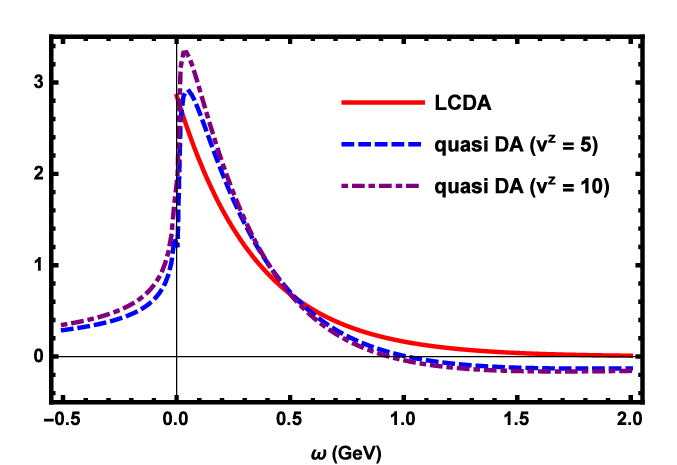}
\includegraphics[width=0.9\columnwidth]{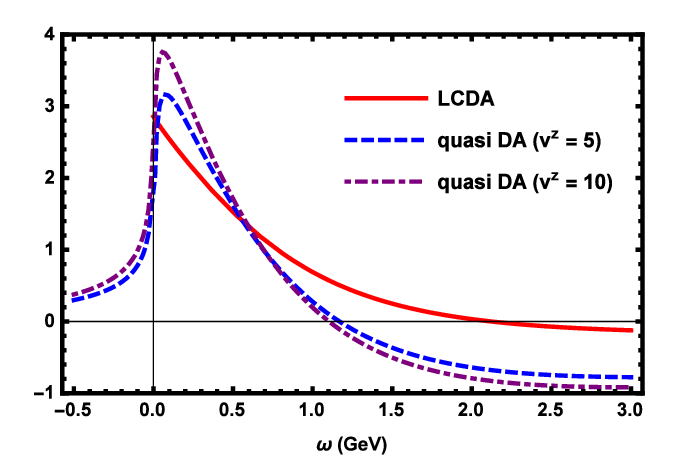}
\centering
\caption{The shapes of the $B$-meson quasidistribution amplitude $\varphi_B^-(\xi=\omega,\mu=1.5\, \textrm{GeV})$ with two different values of $v^z$, obtained from the hard-collinear factorization theorem and the two nonperturbative models $\phi_{B,\textrm{\uppercase\expandafter{\romannumeral1}}}^{-}$ (upper panel) and $\phi_{B,\textrm{\uppercase\expandafter{\romannumeral2}}}^{-}$ (lower panel).
}
\label{diagram12}
\end{figure}

One essential step in accessing the subleading-twist LCDA is to perform the lattice simulation for the quasidistribution amplitude in the moving $B$-meson frame with $v^z \gg 1$.
Although the lattice simulation of $B$-meson quasidistribution amplitude has not been implemented, it is instructive to understand the characteristic feature of $\varphi_B^-(\xi,\mu)$ with nonperturbative models of $\phi_B^-(\xi,\mu)$. Different from the LCDAs of light-meson which can be expanded in terms of Gagenbauer polynomials, the LCDAs of $B$-meson are more difficult to be modeled. We start with two phenomenological models of $\phi_B^-(\xi,\mu)$,
\begin{eqnarray}
  &&\phi_{B,\textrm{\uppercase\expandafter{\romannumeral1}}}^{-}(\omega) = \frac{1}{\lambda_B} e^{-\omega / \lambda_B} \,, \label{models1} \\
  &&\phi_{B,\textrm{\uppercase\expandafter{\romannumeral2}}}^{-}(\omega, \mu) = -\frac{2}{\pi \lambda_B}\left(\frac{\omega \mu}{\omega^2+\mu^2}+\arctan \frac{\omega}{\mu}-\frac{\pi}{2}\right. \nn\\
  &&\left. \quad +\frac{4\left(\sigma_B-1\right)}{\pi^2}\left\{\operatorname{Im}\left[\operatorname{Li}_2 (\frac{i \omega}{\mu} )\right]-\arctan \frac{\omega}{\mu} \ln \frac{\omega}{\mu}\right\}\right) \,.  \label{models2} \nn\\
\end{eqnarray}
Here the reference value of the logarithmic inverse moment $\lambda_B=350 ~\mathrm{MeV}$, $\sigma_B=1.4$ as well as $\mu=1.5 ~\mathrm{GeV}$ are taken \cite{Braun:2003wx,Parkhomenko:2020wau}. The construction of subleading-twist LCDA $\phi_{B,\textrm{\uppercase\expandafter{\romannumeral1}}}^{-}(\omega)$ in Eq.\,(\ref{models1}) is similar in spirit to the simple leading-twist DA proposed in Ref.\,\cite{Grozin:1996pq}, and Eq.\,(\ref{models2}) shows a more complicated model of $\phi_{B,\textrm{\uppercase\expandafter{\romannumeral2}}}^{-}(\omega, \mu)$. Taking advantage of these phenomenological models and the matching coefficient in Eq.\,(\ref{resultofH}), the hard-collinear factorization formula implies the shapes of quasidistribution amplitudes which are displayed in Fig.\,\ref{diagram12}. It is clear that $\varphi_B^-(\xi,\mu)$ develops a radiative tail at large momentum $\xi$ irrespective to the choice of models for $\phi_B^-(\omega,\mu)$. This property was also observed in the study of leading-twist LCDA \cite{Wang:2019msf}. In addition, the shapes of quasidistribution amplitude are close to the LCDA, which indicates the good convergence of perturbation theory.


It should be stressed here that in this work our major objective is to explore the opportunity of accessing the subleading-twist $B$-meson distribution amplitude by simulating the quasidistribution amplitude on the lattice. Therefore, a dedicated lattice calculation of the proposed quasidistribution amplitude is urgently called for. To accomplish this goal, improved methodologies to control both statistical and systematic uncertainties are needed, as well as further development of computing techniques and resources.

\section{Conclusion}
\label{Conclusion}

Within the framework of LaMET, we propose a factorization formula to extract the subleading-twist $B$-meson LCDA from the quasidistribution amplitude and explore its properties. The one-loop matching coefficient which connects these two quantities is derived. These results enable us to further study the simulation of quasidistribution amplitude on the lattice.
In comparison with the previous model constructions, our analysis supplies a guideline for understanding the $B$-meson LCDAs without relying on specific model, offers a new strategy for systematic and detailed studies of $\phi_B^-(\omega,\mu)$. This would present a step forward towards understanding the patterns of subleading corrections and ultimately allow people to increase the accuracy of QCD predictions for heavy meson decays significantly.


Let us conclude by mentioning a couple of further directions to be addressed in future work. First, to further increase the accuracy of our results, the yet unavailable higher-order perturbative correction to the matching coefficient is required. Second, inspecting the nontrivial relations between the two-particle and three-particle $B$-meson LCDAs due to the QCD equations of motion by taking advantage of the forthcoming results from the current strategy is also of substantial significance. Third, one can expand this approach to other physical quantities, such as the LCDA of
heavy baryons and doubly-heavy baryons, etc. In view of the rapid development of LaMET, it may not be unplausible to expect more important progress in the near future.

\section*{Acknowledgements}
S.M.H and J.X. are supported by National Natural Science Foundation of China under Grant No. 12105247, the China Postdoctoral Science Foundation under Grant No. 2021M702957. W.W. is supported in part by Natural Science Foundation of China under grant No. U2032102,
12090064, 12125503, by Natural Science Foundation of Shanghai. S.Z. wishes to thank his parents for their understanding and support during the preparation of this manuscript.

\end{document}